\documentclass[prd,aps,showpacs,showkeys]{revtex4-1}
\usepackage{bm,amsfonts,latexsym,amsmath,amssymb,amsbsy,graphicx}

\newcommand{\bb}{\begin{eqnarray}}
\newcommand{\ee}{\end{eqnarray}}
\newcommand{\ba}{\begin{align}}
\newcommand{\ea}{\end{align}}

\begin{document}

\title{\bf  Bound states of massive fermions in the Aharonov--Bohm-like fields}
\author{V.R. Khalilov}\email{khalilov@phys.msu.ru}
\affiliation{Faculty of Physics, Moscow State University, 119991,
Moscow, Russia}

\begin{abstract}
Bound states of massive fermions in the Aharonov--Bohm-like fields
have analytically been studied. The Hamiltonians with the Aharonov--Bohm-like potentials
are essentially singular and, so, require specification
of a one-parameter self-adjoint extension.  We construct self-adjoint
Dirac Hamiltonians with the Aharonov-–Bohm (AB) potential
in 2+1 dimensions that are specified by  boundary conditions at the origin.
It is of interest that for some range of extension parameter
the AB potential can bind relativistic charged massive fermions.
The bound-state energy is determined by the AB magnetic flux,
depends upon fermion spin and extension parameter; it is
  periodical function of the magnetic flux. We also construct  self-adjoint Hamiltonians
for the so-called Aharonov--Casher (AC) problem, show that
nonrelativistic neutral massive fermions can be bound by
the Aharonov--Casher background, determine the range of extension parameter in which
fermion bound states exist and find their energies as well as wave functions.
\end{abstract}



\maketitle


\section{Introduction}

The quantum Aharonov--Bohm effect \cite{1} is
an important  phenomenon  analyzed
in various physical situations in numerous works
(see e.g., Ref. \cite{AB-review}).
 Considering  an electron travels in a region with the
magnetic flux restricted to a thin solenoid, the electron wave
function may develop a quantum (geometric) phase, which describes the
real behavior of the electrons propagation.
 Thus, the AB vector potential can produce observable
effects because the relative (gauge invariant) phase of the
electron wave function, correlated with a nonvanishing  gauge
vector potential in the domain where the magnetic field vanishes,
depends on the magnetic flux \cite{khu}.

It was observed that the Aharonov--Bohm problem is governed
by  Hamiltonians that are essentially singular and so require
specification of a one-parameter self-adjoint  extension
in order for them to be treated as  self-adjoint
quantum-mechanical operators \cite{ivt,phg,ampw,aw}.
Self-adjoint Hamiltonians are specified
by  boundary conditions at the singular point.

One-parameter self-adjoint extensions
of the Dirac Hamiltonian for the AB problem in
2+1 dimensions were constructed in
\cite{phg,ampw,vrkh}. In \cite{phg}  a formal
solution was constructed, which describes a bound fermion state
in the field of cosmic string.
New great interest to different effects in the  two-dimensional systems has appeared
recently after successful fabrication of graphene (see, \cite{netall,ngpng,kupgc}).
We note while a description of  electron states in the graphene in \cite{ksn,zji,ifh}
were based on the Dirac equation for massless fermions,
work \cite{gkg} has shown that the massive case can also be created.

It seems that  the physical reason for additional specification of
the above Dirac Hamiltonians is also related  to the interaction
between the fermion spin magnetic moment and the source field  \cite{khmam}.
Since the interaction potential is repulsive or attractive
for different signs of spin projection this feature
must be taken into account in the behavior of wave functions at the origin.
The existence of weakly
bound electron states, which can emerge  due to the interaction
between the electron spin magnetic moment and the AB magnetic field
in 3+1 dimensions, was shown in \cite{kh1}.

Fermion bound states can emerge
 in the Aharonov--Casher problem \cite{ahc} of the motion
of a neutral fermion with an anomalous magnetic moment (AMM)
in the electric field of an electrically charged conducting long straight thin thread oriented perpendicularly to the plane of fermion motion resulting from the
interaction between the AMM of the moving fermion and the electric field \cite{safb}.
Authors \cite{safb} argue that such kind of point interaction also appears in several Aharonov--Bohm-like problems \cite{asp1,as0,sa1,fsa1,ako}.

In this paper, we analyze the AB problem taking into
account the fermion spin term in the Dirac
Hamiltonian. We find all self-adjoint Dirac Hamiltonians
as well as their spectra in the Aharonov–-Bohm potential
in 2+1 dimensions using the so-called form asymmetry
method developed in Refs. \cite{vgt,gtv1}.
In particular, expressions for the wave functions and bound state energies
are obtained as functions of the magnetic flux, spin and extension parameters.
By constructing  self-adjoint Hamiltonians for the Aharonov--Casher problem
we show that fermion bound states exist and find their energies as well as wave functions.
We note that the AB and AC scattering problems were studied
in \cite{khmam,khlm1} using corresponding self-adjoint Hamiltonians.

We shall adopt the units where $c=\hbar=1$.

\section{Self-adjoint radial Dirac Hamiltonians in
 an Aharonov--Bohm potential in 2+1 dimensions}

In two spatial dimensions, the Dirac $\gamma^{\mu}$-matrix algebra
is known to be represented  in terms of the
two-dimensional Pauli matrices $\sigma_j$ and the parameter $s=\pm 1$
can be introduced to label two types of fermions  \cite{hoso}
and  is applied to characterize two states
of the fermion spin (spin ``up" and ``down") \cite{crh,khlee}.
Then, the Dirac Hamiltonian
for a fermion of the mass $m$ and charge
$e=-e_0<0$ in an   Aharonov--Bohm
$A_0=0$, $A_r=0$, $A_{\varphi}=B/r$, $r=\sqrt{x^2+y^2}$, $\varphi=\arctan(y/x)$
potential, is
\bb
 H_D=\sigma_1P_2-s\sigma_2P_1+\sigma_3 m,\label{diham}
\ee
where $P_\mu = -i\partial_{\mu} - eA_{\mu}$ is the
generalized fermion momentum operator (a three-vector).
The Hamiltonian (\ref{diham}) should
be defined as a self-adjoint operator in the Hilbert space $\mathfrak H=L^2(\mathbb R^2)$
of square-integrable two-spinors $\Psi({\bf r}), {\bf r}=(x,y)$
with the scalar product
\bb
(\Psi_1,\Psi_2)=\int \Psi_1^{\dagger}({\bf r})\Psi_2({\bf r})d{\bf r},\quad d{\bf r}=dxdy.
\label{scpr}
\ee
The total angular momentum $J\equiv L_z+ s\sigma_3/2$, where $L_z\equiv
-i\partial/\partial\varphi$, commutes with $H_D$, therefore, we can consider
separately in each eigenspace  of the operator $J$
 and the total Hilbert space is a direct orthogonal sum of subspaces of $J$.

In the real (three-dimensional) space, the quantity $B$ characterizes the flux  of the magnetic field
${\bf H}=(0,\,0,\,H)=\nabla\times {\bf A}= B\delta(x)\delta(y)$ through the surface of
infinitely thin (of the radius $R\to 0$) solenoid. Thus, there appears
the interaction potential of the electron spin magnetic moment with the
magnetic field in the form $-s eB \delta(r)/r$, which is singular and
   must influence the behavior of solutions at the origin.
The ``spin'' potential is invariant under the changes  $e\to -e, s\to -s$, and it hence suffices to consider only the case $e=-e_0<0$ and $e_0B\equiv \mu>0$,  $\mu$ is the magnetic flux
$\Phi$ in units of the elementary magnetic flux $\Phi_0\equiv 2\pi/e_0$. Then, the potential is attractive for $s=-1$ and repulsive for $s=1$.
For  cosmic strings  $\Phi=e/Q$, where $Q$ is the Higgs charge \cite{phg,ampw,aw}.

Eigenfunctions of the Hamiltonian (\ref{diham}) are (see, \cite{khlee1})
\bb
 \Psi(t,{\bf r}) = \frac{1}{\sqrt{2\pi r}}
\left( \begin{array}{c}
f_1(r)\\
f_2(r)e^{is\varphi}
\end{array}\right)\exp(-iEt+il\varphi)~, \label{three}
\ee
where $E$ is  the fermion energy, $l$ is an integer.
The wave function $\Psi$ is an eigenfunction of the
operator $J$ with eigenvalue $j=l+s/2$ and
\bb \check h F= EF, \quad F=\left(
\begin{array}{c}
f_1(r)\\
f_2(r)\end{array}\right), \label{radh}\ee
where
\bb
\check h=is\sigma_2\frac{d}{dr}+\sigma_1\frac{l+\mu+s/2}{r}+\sigma_3m,\quad \mu\equiv e_0B
\label{radh0}
\ee
Thus, the problem is reduced to that for the radial Hamiltonian $\check h$
 in the Hilbert space of  doublets $F(r)$ square-integrable on the half-line.

As was shown in \cite{khlee,khlee1} any doublets $F(r)$, $G(r)$ of the Hilbert space
$\mathfrak H=\mathfrak L^2(0,\infty)$ must satisfy
\bb
\lim_{r\to 0} G^{\dagger}(r)i\sigma_2 F(r)=0. \label{bounsym}
\ee
Then, for $\nu=|l+\mu+s/2|\neq n/2, n=1, 2 ,\ldots$  needed linear independent
solutions of (\ref{radh}) are (see, \cite{khlee})
\bb
U_1(r;E)=A(kr)^{1/2}\left(\frac{2m}{k}\right)^{\nu}\Gamma(1/2+\nu)e^{-i\frac{\pi}{4}(1-s)}
\left(\begin{array}{c}
	\sqrt{E+m}J_{\nu-s/2}(kr) \\ \sqrt{E-m}J_{\nu+s/2}(kr)
\end{array}\right)
\label{u1b0}\ee
and
\bb
U_2(r;E)=B(kr)^{1/2}\left(\frac{2m}{k}\right)^{-\nu}\Gamma(1/2-\nu)e^{i\frac{\pi}{4}(1+s)}\left(\begin{array}{c}
	\sqrt{E+m}J_{-\nu+s/2}(kr) \\ -\sqrt{E-m}J_{-\nu-s/2}(kr)
\end{array}\right),
\label{u2b0}\ee
with the asymptotic behavior at $r\to 0$:
$$
U_1(r;E)=(mr)^{\nu}\left(\begin{array}{c}
	1+s \\ 1-s
\end{array}\right){+}O(r^{\nu+1}),\qquad r\rightarrow{0},
$$
$$
U_2(r;E)=(mr)^{-\nu}\left(\begin{array}{c}
	1-s \\ 1+s
\end{array}\right){+}O(r^{-\nu+1}),\qquad r\rightarrow{0},
$$
where $A,B$ are  complex constants, $k=\sqrt{E^2-m^2}$,
and $J_{\mu}(z)$ are the Bessel functions as well as
\bb
V_1(r;E)=U_1(r;E)+\frac{1}{4s\lambda}\omega(E)U_2(r;E),
\label{e40}
\ee
where $\omega(E)={\rm Wr}(U_1,V_1)$ is the Wronskian:
\bb
\omega(E)={\rm Wr}(U_1,V_1)=\frac{\Gamma(2\nu)\Gamma[-\nu+(1-s)/2]}
{\Gamma(-2\nu)\Gamma[\nu+(1-s)/2]}\frac{(2\lambda)^{-2\nu}}{m^{-2\nu}}4s\lambda
\equiv\frac{\tilde{w}(E)}{\Gamma(-2\nu)},
\label{wr00}\ee
where  $\lambda=\sqrt{m^2-E^2}$.
The doublet $V_1$ also can  be represented via the MacDonald functions:
\bb
V_1(r;E)=C(mr)^{1/2}\left(\frac{m}{\lambda}\right)^{\nu-1/2}\frac{2}{\Gamma(1/2-\nu)}\left(\begin{array}{c}
	K_{\nu-s/2}({\lambda}r) \\ sK_{\nu+s/2}({\lambda}r)
\end{array}\right),
\label{grdoub}
\ee
where $C$ is a complex constant.
We note that
\bb
\nu(\pm l, s=1, \mu)=\nu(\pm l+1, s=-1, \mu).
\label{nrel}
\ee
Any doublet of the domain $D(h)$ must satisfy
\bb
 (F^{\dagger}(r)i\sigma_2 F(r))|_{r=0}= (\bar f_1f_2-\bar f_2f_1)|_{r=0} =0. \label{bounsym1}
\ee
$D(h)$ is the space of absolutely continuous doublets $F(r)$
regular at $r=0$ with $hF(r)$ belonging to $\mathfrak L^2(0,\infty)$.

If $\nu>1/2$ there exist only solutions belonging to the continuous spectrum (\ref{u1b0}).
If $0<\nu<1/2$ equation (\ref{bounsym1}) is not  satisfied and its left-hand side
\bb
(\bar f_1f_2-\bar f_2f_1)|_{r=0} = 4s\lambda(\bar{c}_1c_2-\bar{c}_2c_1). \label{bounsym2}
\ee
Therefore the adjoint operator $h^*$ is not symmetric and we need to construct the nontrivial
self-adjoint extensions of the initial symmetric operator $h^0$.
By means of the linear transformation
\bb
c_{1,2}\rightarrow{c_\pm}=c_1\pm{ic_2}
\label{lintr}
\ee
equation (\ref{bounsym2}) is reduced to the quadratic diagonal form
\bb
(\bar f_1f_2-\bar f_2f_1)|_{r=0} = -i4s\lambda(|c_+|^2-|c_{-}|^2)
\label{bounsym3}
\ee
with the inertia indices $(1,1)$, which means that the deficiency indices
of the symmetric operator $h^0$ for $0<\nu<1/2$ are $(1,1)$.
Equation (\ref{bounsym1}) will be satisfied for any $c_-$ related to $c_+$ by
\bb
c_-=e^{i\theta}c_+,\quad 0\leq\theta\leq{2\pi},\quad 0\thicksim{2\pi}.
\label{cc0}
\ee
The angle $\theta$ parameterizes the self-adjoint extensions $h_{\theta}$ of the symmetric operator  $h^0$. These self-adjoint extensions are different for various $\theta$ except for two equivalent cases $\theta=0$ and $\theta=2\pi$.  If we denote $\xi=\tan(\theta/2)$, then
the relation (\ref{cc0}) is equivalent to
\bb
c_2=-\xi{c_1},\quad -\infty\leq\xi=\tan\frac{\theta}{2}\leq+\infty,\quad{-\infty}\thicksim{+\infty}.
\label{ccxi}
\ee
The values of $\xi=\pm\infty$ are equivalent; they imply $c_1=0$ so we can consider only $\xi=\infty$.
Hence, in the range $0<\nu<1/2$ there is one-parameter $U(1)$-family of the operators $h_{\theta}\equiv h_{\xi}$ with the domain  $D_{\xi}$
\bb
h_\xi{:}\left\{\begin{array}{l}
D_\xi=\left\{\begin{array}{l}
F(r):\;F(r)\;\mbox{are absolutely continuous in}(0,\infty),\;F, hF\in{\mathfrak L}^2(0,\infty), \\
F(r)=C\left[(mr)^{\nu}\left(\begin{array}{c}
	1+s \\ 1-s
\end{array}\right)-\xi(mr)^{-\nu}\left(\begin{array}{c}
	1-s \\ 1+s
\end{array}\right)\right],\;{r\rightarrow{0}},\;-\infty<\xi<+\infty,\\
F(r)=C(mr)^{-\nu}\left(\begin{array}{c}
	1-s \\ 1+s
\end{array}\right)+O(r^{1/2}),\;r\rightarrow{0},\;\xi=\infty
\end{array}\right.\\
 h_\xi F=\check hF,
\end{array}\right.
\label{bounab}\ee
where $C$ is a complex constant.
Then
\bb
U_\xi(r;E)=U_1(r;E)-\xi U_2(r;E)
\ee
and
\bb
V_1(r;E)\equiv V_\xi=U_\xi(r;E)+\frac{1}{4s\lambda}\omega_\xi(E)U_2(r;E)
\ee
with
\bb
\omega_\xi(E)={\rm Wr}(U_\xi,V_\xi)=\omega(E)+4s\lambda\xi,
\label{wrn00}
\ee
where $\omega(E)$ is determined by (\ref{wr00}).
For $-\infty<\xi<\infty$, the energy eigenstates (doublets) in the range $|E|\geq m$
are
$$
F(r)=U_1(r;E)-{\xi}U_2(r;E),
$$
where $U_1(r;E)$ and $U_2(r;E)$ are determined by (\ref{u1b0}) and (\ref{u2b0}) with $0<\nu<1$.
The  operator
$h^0$ is not determined as an unique self-adjoint operator and so the additional
specification of its domain, given with the real parameter $\xi$ (the self-adjoint extension parameter)  is required in terms of the self-adjoint boundary conditions.
It is well to note that the self-adjoint boundary conditions  permit
an integrable singularity in the wave functions at origin. Physically, they
show that the probability current density  is equal to zero at the origin.

The spectrum of the radial Hamiltonian is
determined by   (see \cite{vgt,khlee1})
\bb
\frac{d\sigma(E)}{dE}=\frac{1}{\pi}\lim\limits_{\epsilon\rightarrow{0}}{\rm Im}\frac{1}{\omega_{\xi}(E+i\epsilon)},
\label{specfun}\ee
where the generalized function $\omega_{\xi}(E+i\epsilon)$ is obtained by the analytic continuation of the corresponding Wronskian in the complex plane of $E$. It coincides with the corresponding function $\omega(E)$ on the real axis of $E$.
  It can be verified that in the range $|E|>m$
the functions $\omega(E)$  and $\omega_{\xi}(E)$ are
continuous, complex-valued and  not equal to zero for real $E$;
the spectral function $\sigma(E)$ exists and is absolutely continuous.
Thus, the energy spectrum in the range $|E|\geq m$ is continuous.
In the range $|E|<m (-m<E<m)$ the functions $\omega(E)$  and $\omega_{\xi}(E)$ are real and $\lim\limits_{\epsilon\rightarrow{0}}{\omega}_{\xi}^{-1}(E+i\epsilon)$ can be complex only at the points where $\omega_{\xi}(E)=0$ and the energy spectrum  of bound states is determined by roots of this equation. The Wronskians as a function of the complex $E$ have two cuts $(-\infty, -m]$ and $[m,\infty)$ in the complex plane of $E$, so we determine  the first (second) sheet with ${\rm Re}\lambda>0$ (${\rm Re}\lambda<0$) on the real axis of $E$. Real bound states are situated on the first (physical) sheet.

\section{Relativistic bound fermion states in 2+1 dimensions}

 For negative $\xi$  there exists a bound state.
The bound-state energy  $E_\xi(\nu, s)$ is implicitly determined by equation
$\omega_\xi(E)=0$, i.e.
\bb
\frac{\Gamma(2\nu)\Gamma\left(-\nu+(1-s)/2\right)}
{\Gamma(-2\nu)\Gamma\left(\nu+(1-s)/2\right)}\frac{(\lambda)^{-2\nu}}{m^{-2\nu}}=\xi.
\label{levab}
\ee
Let us write  \bb
\mu=[\mu]+\beta\equiv n+\beta, \label{divi}\ee where
$[\mu]\equiv n$ denotes the largest integer $\le \mu$, and $1>\beta\ge 0$.
Hence $n=0, 1, 2, \ldots$ for $\mu>0$ and $n=-1, -2, -3, \ldots$ for $\mu<0$.
Since signs of $e$ and $B$ are fixed it is enough to consider the only
  case $\mu>0$.  One can suppose that a bound state exists due to the interaction
of the fermion spin magnetic moment with AB magnetic field.

We define particle bound states as the states that tend
to the boundary of the  continuous spectrum $E=m$ upon adiabatically slow switching
of the external field (see, for instance \cite{blp,rtkh}).
For $l+n=0, \mu=\beta>0$ the only (particle) bound state  $s=-1$ satisfies
self-adjoint condition (\ref{bounab}). Rewrite
(\ref{levab}) for this case as follows
\bb
\frac{\Gamma(1-2\beta)\Gamma(1/2+\beta)}
{\Gamma(2\beta-1)\Gamma(3/2-\beta)}\left(\frac{m}{\lambda}\right)^{2\beta-1}=\xi, \quad 1/2>\beta>0
\label{lev0}
\ee
and
\bb
\frac{\Gamma(2\beta-1)\Gamma(3/2-\beta)}
{\Gamma(1-2\beta)\Gamma(1/2+\beta)}\left(\frac{m}{\lambda}\right)^{1-2\beta}=\xi, \quad 1>\beta>1/2. \label{lev1}
\ee
It is easily to see that these equations keep for  $l+n=-1, s=1$.
Since $K_{-\gamma}(z)=K_{\gamma}(z)$ it is seen from Eq. (\ref{grdoub}) that bound fermion states with $l+n=0, s=-1$ or $l+n=-1, s=1$ are doublets represented via two MacDonald functions  $K_{1-\beta}(\lambda r)$ and $K_{\beta}(\lambda r)$.

It follows from Eqs. (\ref{lev0}) and (\ref{lev1}) that an adiabatic increase
of the magnetic
flux $\mu$   between the integers $n\to n+1$ lifts an energy level
$E=m\to E=-m$ (see, also \cite{phg}) on the physical sheet ${\rm Re}\lambda>0$) and
$E=-m\to E=m$ on the second (unphysical) sheet ${\rm Re}\lambda<0$).
The second sheet is below the first one. The given bound-state energy
is decreased (increased) $E=m\to E=-m$ for ${\rm Re}\lambda>0$
($E=-m\to E=m$ for ${\rm Re}\lambda<0$) upon adiabatic increase
of the flux $\Phi$  between the integers $n\to n+1$ and is increased (decreased) $E=-m\to E=m$
($E=m\to E=-m$) upon adiabatic increase
of $\Phi$  between  $n+1\to n+2$.  Therefore,
any bound-state energy is a periodic function of the magnetic
flux similar to the case of the fermion motion in the Aharonov-–Bohm
potential along a closed circle \cite{coh}; it is repeated every time
we change $\mu$ by an integer.  It is interesting that the induced current
due to vacuum polarization
in the AB field is finite periodical function of the
magnetic flux  \cite{jmpt}.

For $\xi=-1$ any  curve $E(\beta)$ is symmetric
upon reflection with respect to the point $\beta=1/2, E=0$.
One also can see there exists at $\beta=1/2$ a normalizable state with $E=0$;
for $\xi$ it lies in the middle of the gap $2m$.
The wave function of this (particle) state is
\bb
F(r)=D(mr)^{1/2}\left(\begin{array}{c}
	1 \\ s
\end{array}\right)K_{1/2}(mr).
\label{doubab0}\ee

We give  few comments.

1. In the range of parameters $0>\xi>-\infty$ the constructed self-adjoint Hamiltonians
$h_{\xi}$ have real localized solutions
(fermionic bound state);  physically they exist if additional
potential (in our case, $s\mu\delta({\bf r})$ type) is attractive.

2. We define antiparticle bound states as the states that tend
to the boundary of the  lower continuum $E=-m$ upon adiabatically slow switching
of the external field. Then, we can  treat an antiparticle
as a particle with opposite signs of  $e, s , E$ and we see
that Dirac Hamiltonian (\ref{radh0}) possesses a conjugation symmetry.

Jackiw and Rebbi \cite{rjcr} were observed  that, in a time-inversion, charge conjugation symmetric theory of one-dimensional Dirac fermions interacting with a solitonic background field (the
kink), the effective Hamiltonian possesses a conjugation symmetry.
Because of this symmetry an isolated nondegenerate, charge-self-conjugate,
zero-energy state (zero mode) lying in the middle of the gap $2m$  exists \cite{rjcr,rjsyp,rj} and the vacuum of the model must acquire a half-integer fermionic charge \cite{rjcr}.
In the presence of a vector potential, the Dirac Hamiltonian
 does not exhibit a charge conjugation symmetry
since a charge coupling treats particles and antiparticles differently.
So the existence of fermion states with zero energy  does not
necessarily imply a fractional fermion number \cite{hokh2}.
The presence of a magnetic field breaks time-inversion invariance.

In the considered case, the wave function (a doublet) of antiparticle $F^a$ is related to that of particle $F$ by means of the charge-conjugation operator given by the Pauli matrix
$C=\sigma_3$, i.e. if $F$ is a solution of the Dirac equation (\ref{radh}) with $(l+\mu), s$ and
energy $E$, then $F^a=\sigma_3F^*$ is also a solution of the same equation,
but with $-(l+\mu), -s, -E$.
For $\xi=-1$  the antiparticle energy as a function of $\beta$ is equal
to zero at $\beta=1/2$, and the wave function of antiparticle
state with $E^a=0$ is $F^a(r)=\sigma_3F^*(r)$, where $F(r)$
is determined by (\ref{doubab0}). Therefore,  the AB vector potential
can yield bound states and localized spin-polarized charged zero modes (see, also \cite{hokh2,fmmp}).
Since $F^a(r)$ does not coincide with $F(r)$  the fermionic charge keeps integer.

3. The behavior of the lowest particle energy level
near the upper boundary $E=-m$ of the lower continuum
in the relativistic AB problem differs from the one in the cutoff Coulomb problem.
In the (cut off) Coulomb problem, the lowest electron energy level  can dive into
the lower continuum $[-m,-\infty)$, then turn into resonance  that
can be described as a quasistationary state with ``complex energy''
(directly associated with the creation of  electron-–positron pair)\cite{grrein}(see,
also, \cite{khepj});
 when the bound state pole disappears from the physical sheet
the quasistationary state pole resides on the second  (unphysical) sheet.

We see  there are not particle bound states diving into
the lower continuum, no quasistationary states with ``complex energy'' in the relativistic AB problem (there is not particle creation); also only fermionic bound states with real $E$ can appear
on the second  sheet.

\section{Bound fermion states in the Aharonov--Casher problem}

 The Dirac--Pauli equation for a neutral fermion with the mass $m$, an AMM $M$  in the
form of the Schr\"odinger equation for the case of fermion motion
in an electric field reads \bb i\frac{\partial\Psi}{\partial
t}= H_{DP}\Psi \label{schr}\ee with the Hamiltonian \bb H_{DP}=
\bm{\alpha}\cdot{\bf P}+iM\bm{\gamma}\cdot{\bf E}+\beta m.
\label{reham}\ee Here ${\bf P}=-i\bm{\nabla}$ is the canonical
momentum operator, $\Psi$ is a bispinor, $\gamma^{\mu}=(\gamma^0,\bm{\gamma}), \bm{\alpha}$ are the Dirac matrices  ${\bf E}$ is the electric field strength.

 Introducing the function \bb
\Psi=\Psi_ne^{-imt} \label{nonfun}\ee and representing $\Psi_n$ in
the form
 \bb \Psi_n = \left(
\begin{array}{c}
\phi\\
\chi
\end{array}\right),
\label{spinor}
 \ee
where $\phi$ and $\chi$ are spinors, we obtain an
equation  for the neutral fermion in the electric field of an electrically
charged homogeneous long straight thin  thread directed along the $z$
axis in the nonrelativistic approximation in the form
\bb i\frac{\partial\phi}{\partial t}=\frac{({\bf
P}-{\bf E}\times {\bf M})^2-M^2{\bf E}^2+M\bm{\nabla}\cdot{\bf E}
}{2m}\phi,\label{fineq}\ee where ${\bf
M}=M\bm{\sigma}$, $\bm{\sigma}$ are the Pauli
 matrices and the term $\bm{\nabla}\cdot{\bf E}$ is equal to
 $4\pi$  times the electric field charge density.

In the Aharonov--Casher field configuration \bb E_x=\frac{ax}{r^2},\quad E_y=\frac{ay}{r^2}, \quad
E_z=0,\quad E_r=\frac{a}{r}, \quad E_{\varphi}=0, \label{0one}
 \ee
is the electric field for an electrically charged homogeneous long straight
thin (a zero radius) thread   and $a/2$ is the total surface charge density.
We also assume that the
projection of the fermion momentum on the $z$ axis is equal to zero.
The radial component of the (macroscopic) electric field is determined by the mean surface
charge density as $\bm{\nabla}\cdot{\bf
E}=4\pi \rho$, and the expression
$\rho=a\delta(r)/4\pi r$,
therefore, well approximates $\rho$.
We seek the solutions of (\ref{fineq}) in the polar coordinates in the form
 \bb
 \phi(t, r, \varphi) &=& \exp(-iE_nt)\sum\limits_{l=-\infty}^{\infty}F_l(r)\exp(il\varphi)
\psi, \label{three}
 \ee
where $E_n$ is the particle energy, $l$ is an integer, and $\psi$ is a constant spinor.
The Hamiltonian of a neutral
fermion in the Aharonov-–Casher background  contains only the matrix $\sigma_3$,
and the wave function $\phi$ therefore
depends only on the number $\zeta$ characterizing the conserved spin projection on the $z$ axis, and its eigenvalue $\zeta =\pm 1$ can be substituted for the operator $\sigma_3$ in (\ref{fineq}). After this substitution, the spin part of the wave function $\psi$ becomes inessential,
and we can consider only the scalar coordinate function $\phi$ depending on
$\zeta$ (see, e.g., \cite{ll}). Thus, the radial Dirac--Pauli equation for the neutral fermion with AMM in the electric field of a thread oriented perpendicular to the plane of fermion motion in 3+1 dimensions in the  nonrelativistic approximation coincides to the nonrelativistic equation in the Aharonov–-Bohm problem  and reads \cite{ahc,khmam}
 \bb
h^nF_l(r)=E_nF_l(r),\quad h^n=-\frac{1}{2m}\left(\frac{\partial^2}{\partial
r^2}+\frac{1}{r}\frac{\partial}{\partial r} - \frac{(l+Ma\zeta)^2}{r^2}
-Ma\frac{\delta(r)}{r}\right).  \label{e13}
 \ee
Here $E_n$ is related to $E$ by $E=m+E_n, |E_n|\ll m$.
We also note that analogous singular term ($\sim \delta(r)/r$) also appears in the quadratic Dirac
equation in the AB problem; there it includes spin parameter in the form of an additional delta-function interaction of spin with magnetic field.
The additional term must influence the behavior of solutions at the origin and
it can be taken into account by means of boundary conditions at the point $r=0$.
In the nonrelativistic AC problem the boundary condition (\ref{bounsym1})
can be given by \cite{khmam} (see, also \cite{gstv,khlema})
\bb
 (\bar f'f-\bar ff')|_{r=0} =0,  \label{bounsymn}
\ee
where $f(r)\equiv F_l(r)/\sqrt{r}$ and $\bar f$ is the complex conjugate function $f$.
Here we restrict ourself with considering the case $\gamma=|l+\zeta Ma|<1$ when
bound states can exist.
Then, for each $l$ in the range $0<\gamma<1$ there is one-parameter $U(1)$-family of
self-adjoint Hamiltonians $h_{\xi}^n$ parameterized by (\ref{ccxi}) with the domain  $D_{\xi}^n$
\bb
h_\xi^n{:}\left\{\begin{array}{l}
D_\xi^n=\left\{\begin{array}{l}
f(r), f'(r) \mbox {are absolutely continuous in}(0,\infty); f, h^n_{\xi}f\in{\mathfrak L}^2(0,\infty), \\
f(r)=A[(mr)^{\gamma}-\xi(mr)^{-\gamma}]+ O(r), r\to 0,\quad -\infty<\xi<+\infty,\\
f(r)=A(mr)^{-\gamma}, r\to 0,\quad \xi=\infty
\end{array}\right.\\
 h_\xi^n f=\check h^nf,
\end{array}\right.
\label{bounab1}\ee
where $A$ is a complex constant.
 It is obvious that the function $f(r)$ are the Bessel functions
of the order $\pm \gamma$. Then, calculating  the corresponding Wronskian
we obtain
\bb
\omega(E_n)=\frac{\Gamma(1+\gamma)}
{\Gamma(1-\gamma)}\left(\frac{2m}{\lambda}\right)^{2\gamma},
\label{wr1}\ee
where $\lambda=\sqrt{-2mE_n}$.
By the analytic continuation of (\ref{wr1}) in the complex plane of $E_n$ we obtain
the function $\omega_{\xi}(E_n+i\epsilon)$.
Now the Wronskian as a function of the complex $E_n$
have a cut $(0, \infty)$ in the complex plane of $E_n$
and  the first (second) sheet is determined
${\rm Re}\sqrt{-2mE_n}>0$ (${\rm Re}\sqrt{-2mE_n}<0$). Real bound
states are situated on the first (physical) sheet.

  It can be verified that in the range $E_n>0$
the functions $\omega(E_n)$  and $\omega_{\xi}(E_n)$ are
continuous, complex-valued and  not equal to zero for real $E_n$;
the function $\sigma(E_n)$ exists and is absolutely continuous.
Thus, the energy spectrum in this range  is continuous.
One can show there also exists a bound state (with $E_n<0$)
in the range of parameters  $-\infty<\xi<0)$ for $0<\gamma<1$  and its energy is determined by
\bb
\frac{\Gamma(1+\gamma)}
{\Gamma(1-\gamma)}\left(\sqrt{\frac{-E_n}{2m}}\right)^{-2\gamma}=-\xi.
\label{enern}
\ee
The bound-state energy is the same on the first and second sheets;
it is given by (compare with formula (90) in \cite{safb})
\bb
E_n=-2m\left(-\xi \frac{\Gamma(1-\gamma)}
{\Gamma(1+\gamma)}\right)^{-1/\gamma}.
\label{enerAC}
\ee
The wave function of bound state is
$N\sqrt{mr}K_{\gamma}(\sqrt{-2mE_n}r)$ where  $N$ is a normalization factor.
Since signs of $M$ and $a$ are fixed it is enough to consider the only
  (attractive) case $Ma<0$ and because of bound states exist
for $\gamma<1$ we must have $Ma<-1$. It is seen
 there are bound states with $\zeta=\pm 1$ for $l=0$  and with $\zeta=1(-1)$ for $l=1(-1)$.
Denote $-Ma\equiv c>0$ rewrite (\ref{enerAC}) for these cases
 as follows
\bb
E_n^0=-2m\left(-\xi \frac{\Gamma(1-c)}
{\Gamma(1+c)}\right)^{-1/c},\quad l=0, 0<c<1,
\label{enerAC1}
\ee
\bb
E_n^{\pm 1}=-2m\left(-\xi \frac{\Gamma(c)}
{\Gamma(2-c)}\right)^{1/(c-1)},\quad l=\pm 1, 0<c<1.
\label{enerAC2}
\ee
It is evident that $E_n^0(c)=E_n^{\pm 1}(c=1-b), 1>b>0$.
This means that  an adiabatic increase  of $c$ in the interval $(0,1)$ lifts  the levels
$E_n^0(c)$ on the first (physical) sheet  and
$E_n^{\pm 1}(c)$ on the second (unphysical) sheet in the opposite direction.
The second sheet is below the first one.

Special case $\gamma=0$ can be of some interest (analogous case
was considered in \cite{kh1,gstv} for the nonrelativistic AB problem
in 2+1 dimensions). One can show  that for $|\xi|=\infty$ the energy
spectrum is continuous and nonnegative  as well as for  $-\infty<\xi<0$
there  exists (in addition to the continuous part of the spectrum)
 one negative level
\bb
E_0=-4me^{2(\xi-{\cal C})},
\label{enerAC0}
\ee
where ${\cal C}=0.57721$ is the Euler constant  \cite{GR}.
The wave function of bound state for $\gamma=0$ is
$N\sqrt{mr}K_0(\sqrt{-2mE_0}r)$.

\section{Summary}

By constructing a one-parameter self-adjoint  extension of
the Dirac Hamiltonian with the AB potential in 2+1 dimensions, we have studied
bound states of fermions in this background. It has been shown that for negative
values of extension parameter $\xi$, the spectrum of
self-adjoint Dirac Hamiltonians, in addition to its continuous part,
has one bound level, therefore, the Aharonov--Bohm vector potential can bind
relativistic charged massive fermions in 2+1 dimensions.
The bound-state energy depends upon extension parameter and is
periodical function of the AB magnetic flux. It is of interest that
the AB vector potential can yield localized spin-polarized charged
zero modes.

We  also have studied  the Aharonov-–Casher problem in the context of the
nonrelativistic limit of the Dirac--Pauli equation in 3+1 dimensions. We show
that the  AC background can bind nonrelativistic neutral
massive fermions, determine the range of extension parameter
in which fermion bound states exist and find their energies as well as wave functions.


\begin{thebibliography}{55}

\bibitem{1} Y. Aharonov and D. Bohm, Phys. Rev., {\bf 115},  485 (1959).

\bibitem{AB-review} M. Peshkin and A. Tonomura,
{\sl The Aharonov--Bohm Effect}, (Springer-Verlag, Berlin, 1989).

\bibitem{khu} K. Huang, {\sl Quarks, Leptons, and Gauge Fields} (World
Scientific, Singapore, 1982).


\bibitem{ivt} I.V. Tyutin, {\sl Electron Scattering by a Solenoid},
Preprint of P.N. Lebedev Institute, No 27 (1974), unpublished;
e-print arXiv:quant-ph/0801.2167 v2.

\bibitem{phg} P. de Sousa Gerbert, Phys. Rev.,  {\bf D40}, 1346 (1989).

\bibitem{ampw} M.G. Alford, J. March-Pussel and F. Wilczek,
Nucl.Phys., {\bf B328}, 140 (1989).

\bibitem{aw} M.G. Alford and F. Wilczek, Phys. Rev. Lett., {\bf
62}, 1071 (1989).


\bibitem{vrkh} V.R. Khalilov, Theoretical and Mathematical
Physics, {\bf 163}, 511 (2010).


\bibitem{netall} K. S. Novoselov et al., Science, {\bf 306}, 666 (2004).

\bibitem{ngpng} A.H. Castro Neto, F. Guinea, N.M. Peres, K.S. Novoselov,
and A.K. Geim, Rev. Mod. Phys., {\bf 81}, 109 (2009).

\bibitem{kupgc} V. N. Kotov, B. Uchoa, V. M. Pereira, F. Guinea and A. H. Castro Neto, Rev. Mod. Phys. {\bf 84}, 1067 (2012).

\bibitem{ksn} K.S. Novoselov et al, Nature, {\bf 438}, 197 (2005).

\bibitem{zji} Z. Jiang, Y. Zhang, H.L. Stormer, and P. Kim,
 Phys. Rev. Lett., {\bf 99}, 106802 (2007).



\bibitem{ifh} I.F. Herbut, Phys. Rev. Lett., {\bf 104}, 066404 (2010).









\bibitem{gkg} F. Guinea, M.I. Katsnelson and A.K. Geim,
  Nat. Phys., {\bf 6} 30 (2009).


\bibitem{khmam} V.R. Khalilov, I.V. Mamsurov, Theoretical and Mathematical
Physics, {\bf 161}, 1503 (2009).

\bibitem{kh1} V.R. Khalilov,  Mod. Phys. Lett., {\bf A21}  1647 (2006).










\bibitem{ahc} Y. Aharonov and A. Casher, Phys. Rev. Lett. {\bf 53}, 319
(1984).

\bibitem{safb} E.O. Silva, F.M. Andrade, C. Filgueiras, and H. Belich, Eur. Phys. J., {\bf C73}(4) 2402 (2013).

\bibitem{asp1} F.M. Andrade, E.O. Silva, M. Pereira, Phys. Rev., {\bf D85}(4), 041701(R) (2012).

\bibitem{as0} F.M. Andrade, E.O. Silva, Phys. Lett., {\bf B719}(4-5), 467 (2013).

\bibitem{sa1} E.O. Silva, F.M. Andrade, Europhys. Lett., {\bf 101}(5), 51005 (2013).

\bibitem{fsa1} C. Filgueiras, E.O. Silva, F.M. Andrade, J. Math. Phys.,
{\bf 53}(12), 122106 (2012).

\bibitem{ako} B. Allen, B.S. Kay, A.C. Ottewill, Phys. Rev., {\bf D53}(12),
6829 (1996).

\bibitem{vgt} B.L. Voronov, D.M. Gitman, and I.V. Tyutin,
Theoretical and Mathematical Physics, {\bf 150}, 34 (2007).


\bibitem{gtv1} D.M. Gitman, I.V. Tyutin, and B.L. Voronov,
{\sl Self-adjoint Extensions in Quantum Mechanics}
(Springer Science+Business Media, New York, 2012).

\bibitem{khlm1} V.R. Khalilov, K.-E. Lee, and I.V. Mamsurov,
Mod. Phys. Lett., {\bf A27}, No 5, 1250027 (2012).

\bibitem{hoso} Y. Hosotani, Phys. Lett., {\bf B319}, 332 (1993).

\bibitem{crh} C.R. Hagen, Phys. Rev. Lett., {\bf 64}, 503 (1990).

\bibitem{khlee} V.R. Khalilov and K.-E. Lee, Journ. Phys., {\bf A44}, 205303 (2011).

\bibitem{khlee1} V.R. Khalilov and K.-E. Lee, Mod. Phys. Lett., {\bf A26}, No 12, 865 (2011).

\bibitem{blp} V.B. Berestetzkii, E.M. Lifshitz and L.P. Pitaevskii,
{\sl Quantum Electrodynamics}, 2nd edn. (Pergamon, New York, 1982).

\bibitem{rtkh} V.N. Rodionov, I.M. Ternov, and V.R. Khalilov, ZhETF, {\bf 71}, No 9, 871  (1976).

\bibitem{coh} D. Cohen, {\sl Lecture Notes in Quantum Mechanics}, e-print arXiv:quant-ph/0605180v5 (2013).

\bibitem{jmpt} R. Jackiw, A.I. Milstein, S.-Y. Pi, and I.S.
Terekhov, {\sl Induced Current and Aharonov--Bohm Effect in
Graphene}, e-print arXiv:cond-mat.mes-hall/0904.2046v3, (2009).

\bibitem{rjcr} R. Jackiw, and C. Rebbi, Phys. Rev., {\bf D13}, 3398 (1976).

\bibitem{rjsyp} R. Jackiw, and S.-Y. Pi, Phys. Rev. Lett., {\bf 98}, 266402 (2007).

\bibitem{rj} R. Jackiw, {\sl FRACTIONAL AND MAJORANA FERMIONS:
The Physics of Zero Energy Modes}, e-print arXiv:cond-mat.str-el/1104.4486v1, (2011).

\bibitem{hokh2} C.-L. Ho, and V.R. Khalilov, Phys. Rev.,  D{\bf 63},  027701 (2000).

\bibitem{fmmp} J.M. Fonseca, W.A. Moura-Melo, and A.R. Pereira, {\sl Bound-states
and polarized charged zero modes in three-dimensional topological
insulators induced by a magnetic vortex},e-print arXiv:cond-mat.mes-hall/1210.3100v2, (2013).

\bibitem{grrein} W. Greiner, J. Reinhardt, {\sl Quantum Electrodynamics},
$4^{th}$ ed. (Springer-Verlag, Berlin Heidelberg, 2009).

\bibitem{khepj} V.R. Khalilov, Eur. Phys. J., {\bf C73}(8) 21 (2013).

\bibitem{ll} L.D. Landau and E.M. Lifshitz,
{\sl Quantum Mechanics}, 3rd ed. (Pergamon, New York, 1977).

\bibitem{gstv} D.M. Gitman, A. Smirnov, I.V. Tyutun, and B.L.
Voronov, {\sl Self-adjoint Schr\"odinger and Dirac operators with Aharonov--Bohm and magnetic-solenoid fields}, e-print arXiv:quant-ph/0911.0946v1 (2009).

\bibitem{khlema} V.R. Khalilov, K.-E. Lee, and I.V. Mamsurov,
{\sl Free and bound spin-polarized fermions in the fields of Aharonov--Bohm kind},
e-print arXiv:quant-ph/1002.2826v1 (2010).






\bibitem{GR} I.S. Gradshteyn and I.M. Ryzhik, {\sl Table of Integrals,
 Series, and Products}, $5^{th}$ ed. (Academic Press, San Diego, 1994).


\end{thebibliography}
\end{document}